# Role of Resultant Dipole Moment in Mechanical Dissociation of Biological Complexes

**Maksim Kouza** [1,2,*], **Anirban Banerji** [2,†], **Andrzej Kolinski** [1], **Irina Buhimschi** [3,4] **and Andrzej Kloczkowski** [2,4]

[1] Faculty of Chemistry, University of Warsaw, Pasteura 1, 02-093 Warsaw, Poland; kolinski@chem.uw.edu.pl
[2] Battelle Center for Mathematical Medicine, Nationwide Children's Hospital, Columbus, OH 43215 USA; Anirban.banerji@nationwidechildrens.org (A.B.); andrzej.kloczkowski@nationwidechildrens.org (A.K.)
[3] Department of Pediatrics, The Ohio State University College of Medicine, Columbus, OH 43215, USA; Irina.Buhimschi@nationwidechildrens.org
[*] Correspondence: mkouza@chem.uw.edu.pl; Tel.: +48-22-55-26-364
[†] Deceased 12 August 2015

**Abstract:** Protein-peptide interactions play essential roles in many cellular processes and their structural characterization is the major focus of current experimental and theoretical research. Two decades ago, it was proposed to employ the steered molecular dynamics (SMD) to assess the strength of protein-peptide interactions. The idea behind using SMD simulations is that the mechanical stability can be used as a promising and an efficient alternative to computationally highly demanding estimation of binding affinity. However, mechanical stability defined as a peak in force-extension profile depends on the choice of the pulling direction. Here we propose an uncommon choice of the pulling direction along resultant dipole moment (RDM) vector, which has not been explored in SMD simulations so far. Using explicit solvent all-atom MD simulations, we apply SMD technique to probe mechanical resistance of ligand-receptor system pulled along two different vectors. A novel pulling direction—when ligand unbinds along the RDM vector—results in stronger forces compared to commonly used ligand unbinding along center of masses vector. Our observation that RDM is one of the factors influencing the mechanical stability of protein-peptide complex can be used to improve the ranking of binding affinities by using mechanical stability as an effective scoring function.

## 1. Introduction

Discovery of a new effective drug is a costly and time-consuming process. Billions of US dollars and years in research are spent to place an approved drug on the market. The cost of success is very high due to the fact that many drug candidates fail. One of the possibilities to reduce costs and improve efficiency in current drug discovery processes is to use computer-aided drug design. With the help of molecular modeling, one can predict the success of a potential new drug based on its ability to bind strongly to the target. One of the most popular computational approaches to estimate binding energy is molecular docking simulation by AutoDock (1), whereby the bound conformation of ligand-receptor complex is predicted followed by binding affinity estimation. AutoDock tool can be used for high-throughput virtual drug screening involving thousands to millions of drug candidates. However, it is worth noting that its high performance comes at the cost of accuracy. Limitations of AutoDock and other similar software packages that neglect entropic and solvation effects as well as dynamics properties of the receptor lead to lower accuracy compared to more sophisticated methods such as exact free energy perturbation calculations (2) and molecular mechanics Poisson–Boltzmann surface area (MMPBSA) approach (3). The first method archived unprecedented level of accuracy establishing an astonishing agreement between experimental and computationally predicted values of binding affinities (2). The later approach is an efficient method for the estimation of relative binding affinity for diverse biomolecular systems in reasonable time, however at present applicability of both methods for screening large compounds libraries is limited. Fast and simple methods based on a single or a minimal set of biomolecular structural features, which will be able to reveal latent details in quantitative terms about the strength of protein-peptide complex in a consistent and general manner, are still lacking. Consequently, further development of effective protein-peptide docking techniques (4-7)and finding an efficient alternative to binding affinity (8-15) have been a major focus of computational studies in recent years.

Recently, steered molecular dynamics (SMD) simulations have become popular to measure mechanical stability which could be used to assess the strength of the molecular interactions. The SMD approach was shown to be an efficient alternative to conventional MMPBSA method, but it can be few orders of magnitude faster (9), which enables screening of a correspondingly larger number of compounds. Such gain in performance is possible due to extreme conditions used in SMD simulations, e.g., the pulling speed in simulation is several orders of magnitude higher than that used in single molecule force spectroscopy experiments. Recent studies claim that mechanical unfolding pathways of some proteins are insensitive to pulling forces and speeds if all-atom explicit solvent simulations are employed (16-18). Therefore, it is reasonable to assume that the mechanical stability measured as a force required to unbind a ligand from the receptor corresponds to the strength of interactions. In other words, mechanical stability computed in explicit solvent all-atom SMD simulations could be efficiently used to assess the strength of molecular interactions much faster than conventional methods like MMPBSA.

SMD simulations which mimic the Atomic Force Microscopy (AFM) experiment have been successfully used to study many processes including protein unfolding (19), enzyme-inhibitor unbinding (9) and disaggregation of beta-amyloid oligomers (20). In our previous paper, we demonstrated that kinetic stability of the fibril state can be accessed via mechanical stability extracted from SMD simulation in such a way that the higher mechanical stability or kinetic stability the faster fibril formation (21). A common strategy in SMD simulations applied to single molecules is to pull a protein by force ramped linearly with a time and monitor the mechanical stability as a function of the end-to-end displacement (or time). More than two decades ago, SMD simulations were utilized to measure the interaction strength of the streptavidin-biotin complex. The idea behind using SMD simulations is that the mechanical stability or rupture force required to unbind peptide from the receptor corresponds to the strength of the interactions or in other words peptide mechanical stability is proportional to its binding energy. The ligand was pulled along the vector connecting center of masses (COM) of receptor and ligand. Primarily due to easy implementation of COM's pulling, this direction has become a widely accepted option in MD studies of ligand unbinding. However, it should be pointed out that pulling in the direction connecting the COMs of the protein-peptide system does not necessary align the force vector with it. In this work, we attempt to identify the most prominent non-bonded interaction-based force which may act as a crucial determinant that influences the stability of the protein-peptide complex.

Recently it has been shown (22) that any protein molecule in solution can be represented by a set of polarizable dipoles embedded in a dielectric medium of solvent molecules. Taking a clue from this study, we investigated the role of the resultant dipole moment vector emerging out of the local stretch of protein backbone. In contrast to COMs pulling, the electrostatic force emerging out of the resultant dipole moment ensures the stability of any protein or biological complex. The resultant vector of the peptide-dipoles characterizing the local stretch of protein backbone in the peptide binding site may act as an important determinant of the mechanical strength of protein-peptide complex, especially because, the side-chain dipole moments may either neutralize itself, or, may become neutralized by the innumerable non-bonded interactions which dominate the interactional space involving disordered regions.

In this paper, we investigate the effect of the novel pulling direction on the mechanical stability of ligand-receptor complex using solvent all-atom SMD simulations. As follows from the studies of mechanical unfolding of proteins, the rupture force (or unfolding time in constant force experiments) is sensitive to the pulling direction (23, 24). To our best knowledge, the idea of the pulling ligand from the receptor along resultant dipole moment vector has not been previously explored. For the calmodulin N-lobe bound with ER alpha peptide complex (we will refer to it simply as 2LLO in the rest of the paper) studied here we show that pulling along RDM vector results in stronger forces compared to pulling along COMs vector. We conclude that resultant dipole moment is an important factor influencing the mechanical stability of biological complexes. This can be used to improve the ranking of binding affinities by using mechanical stability or its derivatives as effective scoring functions.

## 2. Results and Discussion

*2.1. Assessing the Mechanical Stability of 2LLO Peptide-Protein Complex Using Steered Molecular Dynamics*

The recent advancement of single-molecule force-spectroscopy (SMFS) techniques has allowed us to detect forces in the pico-newton range (25-28). As a force necessary to unfold protein is in the order of piconewtons, SMFS techniques have become not only one of the most widely applied to study the mechanical unfolding and refolding of biomolecules (25-31) but also a powerful tool to probe the binding of ligand to receptor (9, 10, 12, 32). One of the strategies used in SMFS is to pull a protein by force ramped linearly with time, while monitoring the mechanical resistance as a function of distance between protein ends. The resulting force is computed for each time step to generate a force-extension profile, which has a peak(s) corresponding to the most mechanically stable region(s) in the protein. A typical force-extension profile obtained by constant velocity stretching experiments for a multi-domain construct of the I27 domain is shown in Figure 1a. Each peak of ~200 pN in the force-extension profile arises due to sequential unfolding of the individual domains (25). This remarkable finding was subsequently reproduced by all-atom SMD simulations developed to mimic SMFS experiments (19).

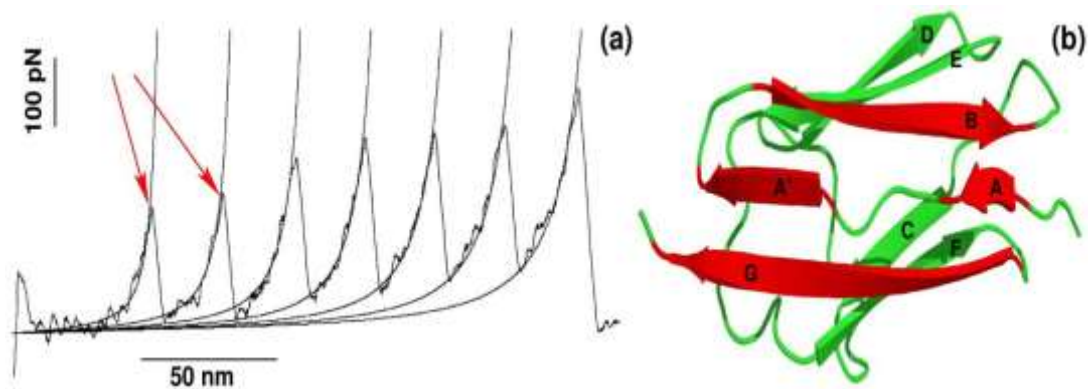

**Figure 1.** (**a**) Dependence of the force as a function of extension for stretching of multi-domain protein titin. The peaks correspond to the unfolding of individual domains with maximum resisting force to stretching, $F_{max}$. Figure adopted from Ref. (25); (**b**) The 3D structure of titin (Brookhaven PDB databank; PDB ID 1TIT). Titin has eight β-strands: A (4–8), A' (11–15), B (18–25), C (32–36), D (47–52), E (55–61), F (69–75), G (78–88). Each peak in force-extension profile corresponds to the breaking of hydrogen bonds between beta-strands marked by red color.

It is worth pointing out that apart from mechanical protein stability measured as $F_{max}$ in the force-extension profile, SMD simulations can be used to investigate the molecular determinants of mechanical stability. Using all-atom explicit solvent SMD simulations (17, 26, 33-35), it was found that each peak in the force-extension profile is associated with breaking hydrogen bonds between strands A and B as well as A' and G in a single domain of a multi-domain construct (Figure 1b). Thus, not only protein mechanical stability, but also molecular interactions and the mechanism behind mechanical unfolding can be revealed using SMD simulations.

Using SMD simulations we undertook a detailed and systematic investigation to quantify the mechanical stability of 2LLO peptide-protein complex. The structure of 2LLO complex has been determined by NMR spectroscopy (36) and its structure in cartoon representation is shown on left of Figure 2 (marked by NS) with alpha-helical ligand colored red and four-helix receptor colored black. We regulated the local environment and applied dissociating force by employing exactly the same set of criteria. We employed the constant pulling speed method for SMD studies. Figure 2 shows the typical force-extension curves for pulling speed $v$ = 0.01 nm/ps for 2LLO system. For the system studied, one distinct peak was consistently observed, which corresponded to the detachment of the peptide from the receptor. Application of a force leads to the external perturbation which drives the system away from equilibrium. At the beginning the force dependence on extension is almost linear obeying the Hooke law. The peak shows the most mechanically stable conformations of protein-ligand complex. Once the hydrogen bonds and van der Waals interactions are broken, the force drops drastically and ligand no longer resists force. The peak in force-time profiles appears to be similar to the two different pulling directions studied, but the height of the peaks is different. Typical conformations observed before and after the occurrence of this peak are shown as snapshots in Figure 2. The separating force was found to drop drastically, though expectedly, once the interactions between peptide and protein ruptured, that is because a ligand can no longer resist the applied force after detachment from the receptor.

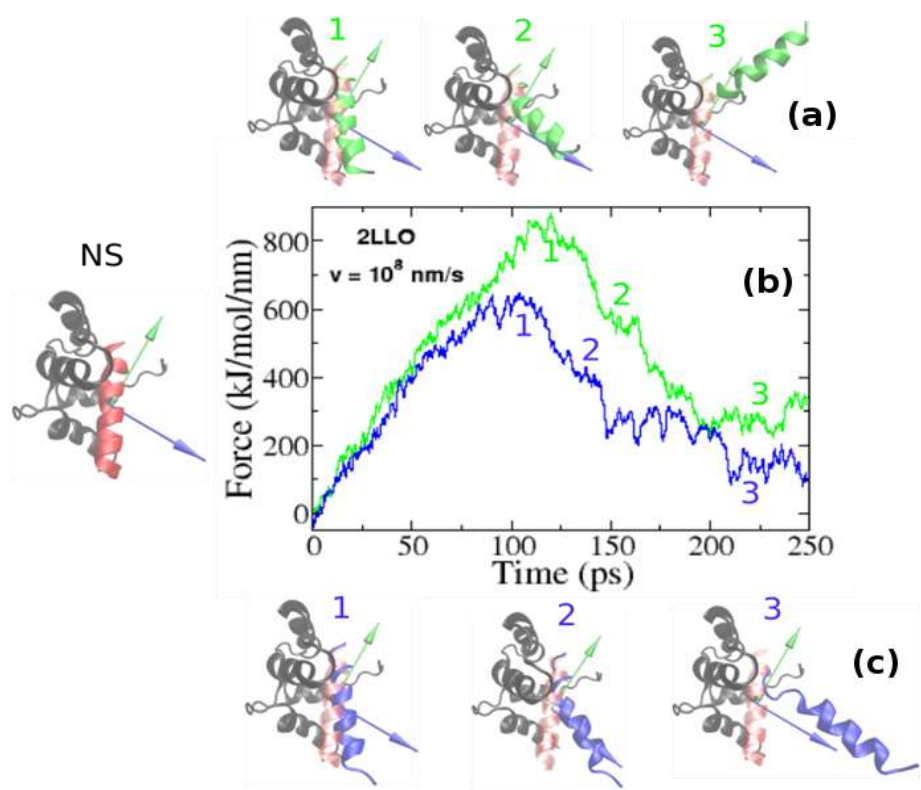

**Figure 2.** Examples of force-extension profiles for 2LLO complex pulled in different directions(b). Green and blue colors refer to the resultant dipole moment (RDM) and COMs pulling directions, respectively. The native conformation of 2LLO is shown on the left (marked by NS) with ligand colored red and receptor colored black. Representative snapshots of pathways for the mechanical unfolding along COMs and RDM directions are shown at the top(a) and bottom(c), respectively. In representative snapshots, we show the position of ligand in native conformation in transparent red.

*2.2. Mechanical Stability Depends on the Pulling Direction of 2LLO Ligand-Receptor Complex*

To elucidate the role of pulling direction on ligand mechanical stability, we have computed mechanical stability, $F_{max}$, for 2LLO protein-peptide complex. Figure 2 shows typical examples of force-extension curves for two different pulling directions and histograms of rupture forces computed from 50 trajectories are presented in Figure 3. The position of the peak corresponding to the most probable rupture force moves toward higher values for the RDM vector compared to the COM vector. The difference between $F_{max}$ for RDM and COM pulling directions is around 180 kJ/mol/nm and indicates that pulling direction alters the mechanical stability of protein-peptide complex drastically. If we consider the averaged value of rupture force, $F_{av}$, we can see that the value of $F_{av}$ is 828 ± 119 and 613 ± 57 kJ/mol/nm for RDM and COM pulling directions, respectively. Thus, regardless of whether the averaged or the most probable rupture force was used, we found that pulling the ligand along the resultant dipole moment vector results in stronger forces compared to the ligand unbinding along the center of masses vector.

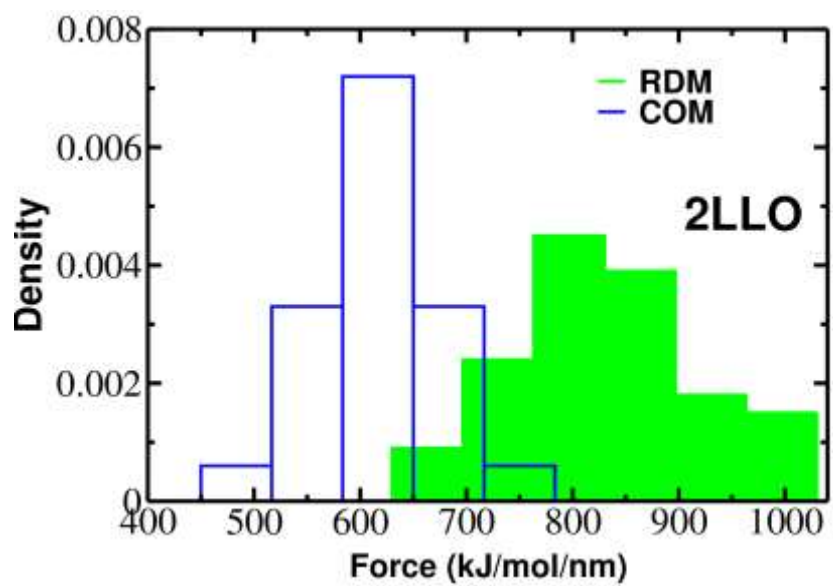

**Figure 3.** Histograms of rupture forces for 2LLO peptide-protein complex along RDM (green) and COM (blue) directions. The histograms clearly show that force peak moves towards higher values for RDM vector compared to COMs one.

It should be noted that the investigation of mechanical dissociation of a biological complex at lower pulling rates by explicit solvent all-atom MD simulations is still a challenge due to the enormous computation time required. However, the knowledge gathered from a two-decade-long spectrum of protein unfolding studies provides further evidence to support the claim that the difference in mechanical stability observed in high force regime will remain robust in low force regime. This seems to be supported by the Bell theory (37), which proposed that the most probable rupture force, $F_{max}$, decreases logarithmically as pulling rate, $v$, is lowered, e.g., $F_{max} \sim ln(v)$. The logarithmic dependence of $F_{max}$ on the pulling speed $v$ was confirmed by numerous experiments and simulations (17, 38, 39).

Our finding shows that pulling a ligand from the receptor along RDM vector results in stronger mechanical stability compared to pulling along COM vector. To make sure that our finding is valid for other protein-peptide systems, in the next subsection, we performed additional simulations on two different protein-peptide complexes.

*2.3. Robustness of Results Against Different Protein-Peptide Complexes*

So far, we have performed SMD simulations for 2LLO protein-peptide complex. The important question arises whether the effect of superior mechanical stability of RDM over COM vectors is universal and holds also for other protein-peptide complexes. In order to check whether our approach holds the same increase on dissociation force trend compared to COM, we performed additional SMD simulations for two different protein-peptide systems. Unlike the first protein-peptide complex that has an alpha-helical peptide, we chose two protein-peptide systems with different classes of peptidic ligands. We used Homer EVH1 domain with bound MGLUR peptide (40) where the bound peptide is unstructured (pdb code is 1DDV) and the inhibitor of apoptosis protein DIAP1 with bound N-terminal peptides from Hid and Grim (41) where the bound peptide takes a β-strand form (pdb code is 1JD5) (See Figure 4a,b). We generated 50 trajectories for each system at pulling rate of 0.01 nm/ns like before. Figure 4a,b show the native conformations of both complexes, while histograms of rupture forces are presented in Figure 4c,d. We computed the values of averaged rupture forces which are presented in Table 1. The difference between different pulling directions for both systems are easily identified, however it should be noted that the difference between RDM and COM for 2LLO is more noticeable compared to 1DDV and 1D5J systems. Intuitively, this can be explained by the difference in peptide size (Table 2). The size of the peptide in 2LLO is nearly 2–3 times larger compared to the peptides in 1DDV and 1D5J protein-peptide systems. Overall, our finding demonstrates that pulling a peptidic ligand from the receptor along RDM vector results in stronger mechanical stability compared to the pulling along COM for three diverse peptide-protein systems. Thus, regardless of the protein-peptide system used, pulling along RDM vector results in higher mechanical stability compared to the commonly used COM pulling.

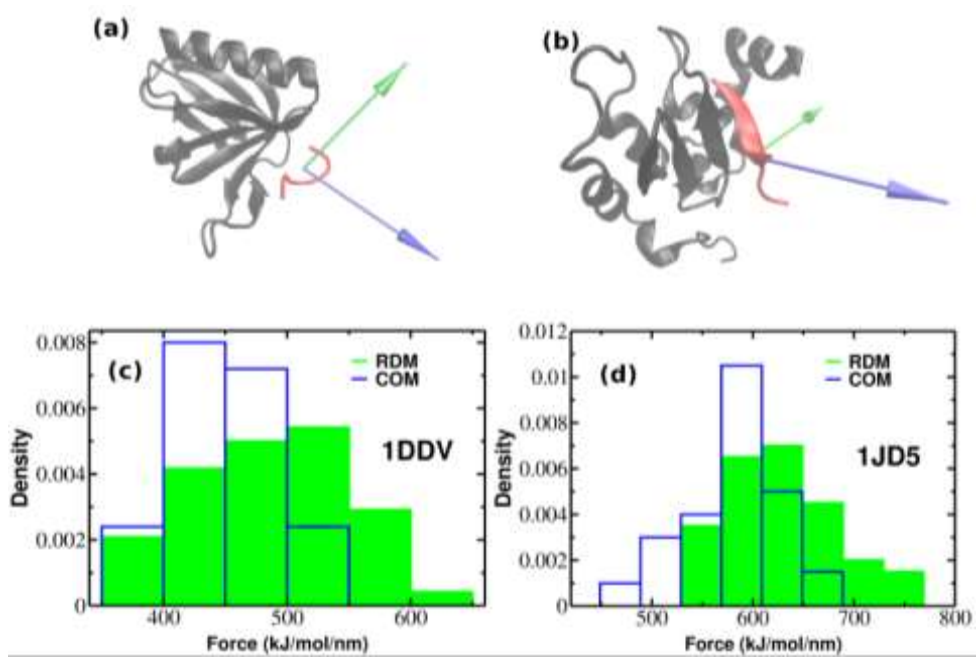

**Figure 4.** The native conformations of 1DDV (**a**) and 1JD5 (**b**) complexes with ligand colored red and receptor colored black. Green and blue colors refer to the resultant dipole moment (RDM) and COMs pulling directions, respectively. Histograms of rupture forces for 1DDV (**c**) and 1JD5 (**d**) peptide-protein complexes along RDM (green) and COMs (blue) directions.

**3. Materials and Methods**

We used GROMOS43a1 (42) force-field (43) to describe the peptides and SPC (44) water model for solvent. All-atom MD simulations have been carried out using Gromacs program suite (45) which was previously successfully employed by our group for studying protein folding, unfolding and aggregation (46-49). We use periodic boundary conditions and calculate the electrostatic interactions by the particle mesh Ewald method (50). The non-bonded interaction pair-lists are updated every 10 fs, using a cutoff of 1.4 nm. All bond lengths are constrained with the linear constraint solver LINCS (51), allowing us to integrate the equations of motion with a time step of 2 fs.

To avoid improper structures, the whole system was minimized with the steepest-descent method, before being equilibrated at 310 K with two successive molecular dynamics runs of length 1ns each; the first one at constant volume, the second at constant pressure (1 atm). Initial velocities of the atoms were generated from the Maxwell distribution at 310 K. The temperature was kept close to 310 K using the v-rescale thermostat. Data analysis was done using the corresponding Gromacs programs and snapshots of all peptides were created with Visual Molecular Dynamics molecular graphics software (52). Resultant dipole moment was defined as net dipole moment of those receptor backbone atoms which interacts with the bonded ligand.

During the steered molecular dynamics (SMD) simulations, the spring constant was chosen as $k = 1000$ kJ/(mol·nm²) ≈ 1700 pN/nm, which corresponds to the upper limit of $k$ of cantilever used in AFM experiments. We applied an external force to the center of mass (COM) of the ligand and pulled it along two different vectors. The first vector is drawn between COM of pulled peptide and COM of the receptor. The second vector is the resultant dipole moment defined as net dipole moment of those receptor backbone atoms which interacts with the bonded ligand. Pulled movement of the peptide under external force caused its dissociation from the receptor and the total force needed to bring about this dissociation was measured by $F = k(vt - x)$, where $x$ denoted the displacement of the pulled peptide from its initial position. The resulting force was computed for each time step to generate a force-extension profile, which recorded a single peak showing the most mechanically resisting conformation in our system. Once the critical interactions were disrupted, the pulled peptide was found to no longer resist the applied force. Overall, the simulation procedure could be described similarly to those followed during the AFM experiments, except that the pulling speeds in our SMD simulations were fixed at several orders of magnitude higher than those used in AFM experiments (53). We performed simulations at room temperature (T = 310 K) for $v = 10^7$ nm/s and generated 50 trajectories for each pulling direction. The 50 peak forces extracted were subsequently used to construct histogram of most probable rupture forces.

**Table 1.** List of three protein-peptide complexes used in all-atom SMD simulations. Amino acids of protein involved in protein-peptide interactions for each system are shown. The pulling vectors used in the all-atom simulations (RDM and COM) and the averaged rupture forces and standard deviations obtained along RDM and COM pulling are also shown. Data are averaged over 50 trajectories.

| PDB Code of the Protein-Peptide Complex | Identified Protein Residues Involved in Protein-Peptide Interactions | RDM Vector | COM Vector | Force (kJ/mol/nm) RDM | COM |
|---|---|---|---|---|---|
| 2LLO | 7–21, 25–29, 31–40, 43, 47–58, 61–65, 67–80 | $0.144i + 0.983j - 0.11k$ | $0.193i + 0.068j + 0.979k$ | 828.5 ± 118.7 | 613.5 ± 56.4 |
| 1DDV | 10–16, 22–26, 30–31, 69–76, 87–92, 96, 109 | $-0.636i + 0.111j + 0.764k$ | $-0.799i + 0.454j - 0.395k$ | 486.1 ± 53.6 | 432.4 ± 42.1 |
| 1JD5 | 219–220, 242, 252–257, 259–263, 265–279, 282–283, 285–290, 311, 314–315, 317–318 | $0.226i + 0.182j - 0.957k$ | $-0.77i - 0.416j - 0.484k$ | 773.9 ± 149.4 | 595.9 ± 55.1 |

**Table 2.** List of 3 complexes used in the all atom SMD simulations. The bound structures of the complexes are obtained from the structures deposited in Protein Data Bank. The lengths and structural classes of both the proteins and the peptides are provided.

| PDB ID of The Protein-Peptide Complex | Protein | | Peptide | |
|---|---|---|---|---|
| | Length | Class | Length | Class |
| 2LLO | 80 | α/β (34/2%) | 19 | α (84%) |
| 1DDV | 104 | α/β (13/45%) | 6 | unstructured |
| 1JD5 | 105 | α/β (41/7%) | 8 | β (40%) |

## 4. Conclusions

In the reported here studies we have tested the influence of RDM pulling direction on mechanical stability of three peptide-protein complexes. Unlike in widely-used COMs pulling simulations where COM does not talk about the forces that contribute to the stability of the complex, RDM vector retains information about the electrostatic forces associated with the resultant dipole moment. Pulling along COMs vector turns out to lead to a weaker resistance compared to RDM direction which has a significant electrostatic force aligned with it. Thus, together with other geometric and dynamics properties of protein binding pockets (54), RDM is one of the important factors influencing stability of biological complexes. Consequently, we hypothesize that peptide ligand binding affinity might be more accurately predicted using mechanical stability obtained by a computational approach that incorporates RDM factor in SMD studies. Our finding can provide a basis, through qualitative, for improvement of the computationally predicted mechanical stability. We believe that this should lead to development of new strategies that employ the mechanical stability as an effective scoring function for ranking binding affinities and/or for the quick testing of peptide ligands that might eventually block formation of pathological aggregates.

**Author Contributions:** M.K. conceived research, M.K. and A.B. performed simulations and analyzed data, and all authors wrote manuscript.

**Acknowledgments:** The second author, Anirban Banerji, passed away in Columbus, OH, USA on 12 August 2015 at the age of 39. This research was supported in part by the High Performance Computing Facility at The Research Institute at Nationwide Children's Hospital. A. Kol. and M.K. would like to acknowledge support from the National Science Center grant [MAESTRO 2014/14/A/ST6/00088]. M.K. acknowledges the Polish Ministry of Science and Higher Education for financial support through ''Mobilnosc Plus'' Program No. 1287/MOB/IV/2015/0. IAB acknowledges support from the Eunice Kennedy Shriver National Institute of Child Health and Human Development (NICHD) R01HD084628 and The Research Institute at Nationwide Children's Hospital's John E. Fisher Endowed Chair for Neonatal and Perinatal Research. A Klo acknowledges support from NSF (DBI 1661391), NIH R01GM127701 and The Research Institute at Nationwide Children's Hospital.

**Conflicts of Interest:** The authors declare no conflict of interest.